\documentclass[aps, twocolumn, superscriptaddress, longbibliography]{revtex4-2}
\usepackage[english]{babel}
\usepackage{graphicx}
\usepackage{amsmath, amsfonts, amssymb}
\usepackage{amsthm}
\usepackage{mathtools}
\usepackage{bm}

\usepackage{bbm}
\usepackage{tikz}
\usetikzlibrary{positioning,chains,fit,shapes,calc}
\usepackage[
colorlinks=true,
urlcolor=blue,
citecolor=blue,
linkcolor=blue,
hyperfootnotes=false]{hyperref}
\usepackage{cleveref}
\usepackage[mode=buildnew]{standalone}%

\DeclareMathOperator{\im}{im}
\DeclareMathOperator{\rank}{rk}

\begin{document}
 \title{Improved single-shot decoding of higher dimensional hypergraph product codes}
	
	\author{Oscar Higgott}
	\email{oscar.higgott.18@ucl.ac.uk}
	\affiliation{Department of Physics \& Astronomy, University College London, WC1E 6BT London, United Kingdom}
	\author{Nikolas P. Breuckmann}
	\email{n.breuckmann@ucl.ac.uk}
	\affiliation{Department of Computer Science, University College London, WC1E 6BT London, United Kingdom}
	\date{\today}
	
	\begin{abstract}
    In this work we study the single-shot performance of higher dimensional hypergraph product codes decoded using belief-propagation and ordered-statistics decoding~\cite{Panteleev2021degeneratequantum}.
    We find that decoding data qubit and syndrome measurement errors together in a single stage leads to single-shot thresholds that greatly exceed all previously observed single-shot thresholds for these codes.
    For the 3D toric code and a phenomenological noise model, our results are consistent with a sustainable threshold of 7.1\% for $Z$ errors, compared to the threshold of 2.90\% previously found using a two-stage decoder~[Quintavalle et al., 2021].
    For the 4D toric code, for which both $X$ and $Z$ error correction is single-shot, our results are consistent with a sustainable single-shot threshold of 4.3\% which is even higher than the threshold of 2.93\% for the 2D toric code for the same noise model but using $L$ rounds of stabiliser measurement.
    We also explore the performance of balanced product and 4D hypergraph product codes which we show lead to a reduction in qubit overhead compared the surface code for phenomenological error rates as high as 1\%.
	\end{abstract}

	\maketitle
	
\section{Introduction}

Quantum error correction will be necessary to protect quantum computers from errors due to unwanted interactions with the environment.
One of the most promising error correcting codes is the surface code, which can be implemented using stabiliser measurements that are geometrically local in two-dimensional planar architectures, and has a high threshold for realistic noise models.
However, for the surface code, stabiliser measurements must be repeated many times to handle the noise that is inevitably present in the stabiliser measurements themselves.
Furthermore, one surface code patch only encodes a single logical qubit ($k=1$) in $n$ physical qubits, leading to a vanishing encoding rate $k/n$ as the block length $n$ is increased~\cite{dennis2002topological}.
These limitations lead to large time and space (qubit) overheads, with daunting resource requirements for realising fault-tolerant quantum computation~\cite{fowler2012surface}.

The time overhead for fault-tolerant quantum computation can be significantly reduced by instead using \textit{single-shot} quantum error correction, introduced by \citeauthor{bombin2015single}~\cite{bombin2015single}.
With single-shot quantum error correction, only a single round of stabiliser measurements is sufficient when decoding, even in the presence of measurement errors.
Codes that have been shown to support single-shot error correction include the 3D gauge color code~\cite{bombin2015single,brown2016fault}, 3D toric and 3D hypergraph product codes (for $Z$ noise)~\cite{kubica2019cellular, vasmer2021cellular,quintavalle2020single}, 4D toric and 4D hypergraph product codes~\cite{breuckmann2016local,campbell2019theory,zeng2019higher}, 4D hyperbolic codes~\cite{breuckmann2020single}, 3D subsystem toric codes~\cite{kubica2021single} and quantum expander codes~\cite{fawzi2020constant}.
More generally, \citeauthor{quintavalle2020single} defined a property called \textit{confinement}, and showed that codes satisfying this property are single-shot under an adversarial noise model, with linear confinement sufficient for single-shot error correction under a stochastic noise model~\cite{quintavalle2020single}.
Of particular interest are single-shot codes that also have improved code parameters relative to the surface code~\cite{campbell2019theory,zeng2019higher,breuckmann2020single,fawzi2020constant,quintavalle2020single} such as hypergraph product codes~\cite{tillich2013quantum,zeng2019higher,campbell2019theory}, which have a finite encoding rate as well as a minimum distance growing as $O(\sqrt{n})$.

In this work, we focus on the problem of single-shot decoding of 3D and 4D toric codes and 4D hypergraph product codes.
We find that decoding data qubit and syndrome measurement errors together in a single stage using belief propagation and ordered statistics decoding (BP+OSD) leads to very high thresholds.
Using our decoder, we find a threshold of 7.1\% for the 3D toric code with a phenomenological noise model, far exceeding the highest previously reported threshold of 2.90\% found using a two-stage decoder, where syndrome measurement errors are repaired first using minimum-weight perfect matching (MWPM), before correcting data qubit errors using BP+OSD~\cite{quintavalle2020single}.
Another approach for single-shot decoding is using a cellular automaton, the sweep rule, and this approach was found to have a threshold of 1.7\% for the 3D toric code~\cite{kubica2019cellular, vasmer2021cellular}.

Previous work on single-shot decoding of the 4D toric code has mainly considered local decoders, including cellular automaton decoders, as well as a decoder due to Hastings~\cite{hastings2013decoding}, which was found in Ref.~\cite{breuckmann2016local} to have a threshold of 1.59\% for the 4D toric code with a phenomenological noise model.
We instead find a threshold of 4.3\% using single stage BP+OSD decoding which, remarkably, even exceeds the threshold of 2.93\% for the 2D toric code with the same noise model but using $L$ rounds of stabiliser measurements~\cite{wang2003confinement}.

\section{Stabiliser codes}

A stabiliser code is defined as the $+1$-eigenspace of elements of its stabiliser group $\mathcal{S}$, which is an abelian subgroup of the Pauli group $\mathcal{P}_n$ on $n$ qubits which does not contain the element $-I$.
The centraliser $C(\mathcal{S})$ of the stabiliser group is the set of Pauli operators that commute with every element of $\mathcal{S}$.
The weight $|P|$ of a Pauli operator $P\in\mathcal{P}_n$ is the number of qubits on which it acts nontrivially.
Elements of $C(\mathcal{S})\setminus \mathcal{S}$ are nontrivial logical operators, and the minimum distance of the stabiliser code is the minimum weight of any Pauli operator in $C(\mathcal{S})\setminus \mathcal{S}$.

We perform error correction by measuring a set of generators $\{s_0, s_1,\ldots, s_{r-1}\}$ of the stabiliser group $\mathcal{S}$, and we call this set of generators the \textit{check operators}. 
Given an error $E\in\mathcal{P}_n$ and a choice of check operators, the syndrome map $\sigma(E)$ gives the measurement outcome of each check operator.
For a given stabiliser code, the \textit{reduced weight} $|P|^{\mathrm{red}}$ of a Pauli operator $P\in\mathcal{P}_n$ is the minimum weight $|Q|$ of any Pauli $Q\in\mathcal{P}_n$ with the same syndrome $\sigma(P)=\sigma(Q)$.
Since each Pauli operator can be factorised as the product of a $Z$-type Pauli operator in $\{I,Z\}^n$ (which we call its $Z$ component) and an $X$-type Pauli operator in $\{I,X\}^n$ (which we call its $X$ component), we can represent any Pauli operator $P$ (up to phases) as a binary vector $\mathbf{p}=(\mathbf{p}_Z,\mathbf{p}_X)\in\mathbb{F}_2^{2n}$, where the $i$th element of $\mathbf{p}_Z$ is nonzero only if the $Z$ component of $P$ acts nontrivially on the $i$th qubit, and likewise the $i$th component of $\mathbf{p}_X$ is nonzero only if the $X$ component of $P$ acts nontrivially on the $i$th qubit.
Multiplication of Pauli operators is equivalent to addition (modulo 2) of their binary representations, and we can represent the stabiliser group of a stabiliser code as a parity check matrix $H=(H_Z, H_X)$, where the $i$th row of $H$ is the binary representation of the $i$th generator of $\mathcal{S}$.
Given an error $E\in\mathcal{P}_n$ with binary representation $\mathbf{e}$, the binary representation of its syndrome $\mathbf{s}$ is given by the matrix vector product $\mathbf{s}=H\mathbf{e}$.

A Calderbank-Shor-Steane (CSS) stabiliser code has a stabiliser group that can be generated by check operators that are either $X$-type \textit{or} $Z$-type Pauli operators. As a result, its parity check matrix can be written in the form
\begin{equation}
H = 
\left (
\begin{array}{cc}
0 & H_X \\
H_Z & 0
\end{array}
\right )
\end{equation}
where $H_Z$ and $H_X$ each have $n$ columns and are called the $Z$ and $X$ parity check matrices, respectively. The requirement that the stabiliser group be abelian can now be expressed as the condition 
\begin{equation}\label{eq:commutativity}
H_ZH_X^T=0.
\end{equation}

\section{Chain complexes and $\mathbb{F}_2$-homology}

We will now introduce chain complexes and $\mathbb{F}_2$-homology, which we will see later are useful tools for the study of CSS codes. 
For a more comprehensive introduction to chain complexes, and their use in constructions of quantum LDPC codes, we refer the reader to Refs.~\cite{hatcher2002algebraic,breuckmann2021ldpc}.
In $\mathbb{F}_2$-homology, a chain complex $C$ with length $l$ is a collection of vector spaces $C_i\coloneqq\mathbb{F}_2^{n_i}$
$$
\{0\} 
\overset{\partial_{l+1}}{\longrightarrow} 
C_l 
\overset{\partial_l}{\longrightarrow} 
C_{l-1} 
\cdots
\overset{\partial_1}{\longrightarrow} 
C_0
\overset{\partial_0}{\longrightarrow} 
\{0\}
$$
and \textit{boundary maps} $\partial_i:C_i\rightarrow C_{i-1}$, where the boundary maps satisfy the constraint 
\begin{equation}\label{eq:bound2}
\partial_{i}\partial_{i+1}=0
\end{equation}
for all $i\in\{0\ldots l\}$. We refer to each element $C_i$ as an $i$-cell, and call elements of $\im\partial_{i+1}$ and $\ker \partial_i$ \textit{boundaries} and \textit{cycles}, respectively. From~\Cref{eq:bound2} we know that $\im\partial_{i+1}\subseteq \ker \partial_i$ (every boundary is a cycle). However, each cycle is not necessarily a boundary, and the $i$th \textit{homology group} of $C$ is the quotient
\begin{equation}
\operatorname{H}_i=\ker\partial_i/\im\partial_{i+1}.
\end{equation}

Associated with $C$ is another chain complex called a \textit{cochain} complex with \textit{coboundary} operators $\delta^i:C_i\rightarrow C_{i+1}$ defined as $\delta^i\coloneqq \partial_{i+1}^T$:
$$
\{0\} 
\overset{\delta^{-1}}{\longrightarrow} 
C_0
\overset{\delta^{0}}{\longrightarrow} 
C_{1} 
\cdots
\overset{\delta^{l-1}}{\longrightarrow} 
C_l
\overset{\delta^l}{\longrightarrow} 
\{0\}.
$$
Elements of  $\ker \delta^i$ and $\im\delta^{i-1}$ are \textit{cocycles} and \textit{coboundaries} respectively, and the $i$th \textit{cohomology} group is $\operatorname{H}^i=\ker \delta^i / \im\delta^{i-1}$.

\section{Hypergraph product codes}

The commutativity condition for CSS codes, \Cref{eq:commutativity}, is equivalent to the defining property of chain complexes in $\mathbb{F}_2$-homology, given in \Cref{eq:bound2}.
A consequence of this is that we can use a chain complex with length at least two to define a CSS code. We associate each qubit with an $i$-cell, where $0<i<l$. We then use the $i$th boundary operator as the $X$ check matrix $H_X=\partial_i$ and use the $i$th coboundary operator as the $Z$ check matrix $H_Z=\delta^i=\partial_{i+1}^T$. The commutativity condition $H_XH_Z^T=0$ is then guaranteed to be satisfied by \Cref{eq:bound2}. The $Z$ logicals are associated with elements of the $i$th homology group $\operatorname{H}_i$, and the $X$ logicals are associated with elements of the $i$th cohomology group $\operatorname{H}^i$. The number of logical qubits is therefore given by $\dim \operatorname{H}_i=\dim \operatorname{H}^i$.

One method of constructing chain complexes of length at least 2 is by taking the tensor product of length-1 chain complexes.
The tensor product $B\otimes C$ of two complexes $B$ and $C$, with lengths $l_B$ and $l_C$ respectively, is a chain complex $A$ with length $l_A=l_B+l_C$. This product complex $A$ is a collection of vector spaces $A_0, A_1,\ldots,A_{l_A}$ for which
\begin{equation}
A_i=\bigoplus_{j+k=i} B_j\otimes C_{k}
\end{equation}
and where the action of a boundary operator $\partial^A$ in $A$ on vectors $b\otimes c\in B_i\otimes C_j$ is defined to be
\begin{equation}
\partial^A(b\otimes c)=\partial^B_i b \otimes c + b \otimes \partial^C_j c.
\end{equation}
Here $\partial^B_i$ and $\partial^C_j$ are boundary operators acting in $B$ and $C$ respectively. 

The homology groups of the product complex are given by the K\"unneth theorem
\begin{equation}
\operatorname{H}_i(A) \cong \bigoplus_{j+k=i} \operatorname{H}_j(B)\otimes \operatorname{H}_k(C)
\end{equation}
and therefore the rank $r_i^A$ of the $i$th homology group of $A$ is given by
\begin{equation}
r_i^A= \sum_{j+k=i}r_j^Br_k^C.
\end{equation}

The parity check matrix $H$ of a linear code can be considered the boundary map $\partial_1$ of a length-one chain complex $C(H)$. 
Given parity check matrices $H_a$, $H_b$ of two linear codes, we can construct the \textit{hypergraph product code} $HGP(H_a,H_b)$ by taking the tensor product $C(H_a)\otimes C(H_b^T)$ to obtain a length-two chain complex, with which we associate qubits with 1-cells, $Z$-stabilisers with 2-cells and $X$-stabilisers with 0-cells. 
Using linear rate and linear distance LDPC codes for $H_a$ and $H_b$, \citeauthor{tillich2013quantum} showed that the quantum LDPC code $HGP(H_a,H_b)$ on $n$ qubits has linear rate and distance $\Omega(\sqrt{n})$~\cite{tillich2013quantum}.
Specifically, if the check matrices $H_a$ and $H_b$ have dimensions $n_a^T\times n_a$ and $n_b^T\times n_b$ respectively then $HGP(H_a,H_b)$ will have $n_an_b+n_a^T n_b^T$ physical qubits. 
We will denote the nullity of $H_a$, $H_b$, $H_a^T$ and $H_b^T$ by $k_a$, $k_b$, $k_a^T$ and $k_b^T$, respectively.
From the K\"unneth theorem, the number of logical qubits in $HGP(H_a,H_b)$ is $k_1k_2+k_1^T k_2^T$.
We will denote the distance of the codes $\ker H_a$ and $\ker H_b$ by $d_a$ and $d_b$, respectively.
The distance of $HGP(H_a,H_b)$ is then given by $\min(d_a, d_b)$ if $k_a^T=0$ or $k_b^T=0$ or $\min(d_a, d_b,d_a^T,d_b^T)$ otherwise.
The 2D toric code is the hypergraph product of two repetition codes, and as a result hypergraph product codes can be seen as a generalisation of the toric code.

We can construct higher-dimensional hypergraph product codes by taking the product of more than two chain complexes. Taking the hypergraph product of $D$ repetition code chain complexes we obtain a complex~\cite{zeng2019higher}
\begin{equation}
C_{toric}^D\coloneqq C(H_{rep})^{\otimes D}
\end{equation}
from which a $D$-dimensional toric code can be derived. A $D$-dimensional toric code $T_i^D$ is defined by associating a qubit with each $i$-cell of $C_{toric}^D$, a $Z$-stabiliser with $(i+1)$-cell, and an $X$-stabiliser with each $(i-1)$-cell. Hence, we can define $D-1$ unique toric codes for each $C_{toric}^D$, one for each $i$ satisfying $1\leq i\leq D-1$. The $D$-dimensional toric code $T_i^D$ encodes ${D \choose i}$ logical qubits in ${D \choose i}L^D$ physical qubits, with a $Z$ distance of $L^i$ and an $X$ distance of $L^{D-i}$.

We define the \textit{distance} $d_i(C)$ of the $i$th homology group of a chain complex $C$ to be the minimum hamming weight of any nontrivial element of the group, and define its distance to be infinite if the homology group has zero rank. 
In Ref.~\cite{zeng2019higher}, it was shown that the distance $d_i(A\otimes B)$ of the $i$th homology group of the product of a complex $A$ (of any length) with a length-one complex $B$ is given by
\begin{equation}\label{eq:product_distance}
d_i(A\otimes B)=\min(d_i(A)d_0(B), d_{i-1}(A)d_1(B)).
\end{equation}
Consider the chain complex obtained by taking the product
\begin{equation}
C^{a,b}=C(H)^{\otimes a}\otimes C(H^T)^{\otimes b}
\end{equation}
where $H$ is a full rank $r\times c$ parity check matrix of a linear code $
\ker{H}$.
The $a$th homology group of $C^{a,b}$ is the only nontrivial homology group, with rank $\kappa^{a+b}$, where $\kappa$ is the rank of $H$~\cite{zeng2019higher}.
If the distance of the code $\ker{H}$ is $d$, then a direct application of \Cref{eq:product_distance} can be used to show that the distance of a CSS code defined with qubits on $a$-cells of $C^{a,b}$ is $\min(d^a,d^b)$~\cite{zeng2019higher}.
Furthermore, the dimension $n_a(C^{a,b})$ of the $a$th vector space of the complex $C^{a,b}$ is given by 
\begin{equation}
n_a(C^{a,b})=\sum_{i=0}^ac^{b-a+2i}r^{2a-2i}{a\choose i}{b\choose a-i}.
\end{equation}
Therefore the quantum code derived from $C^{a,b}$ by associating qubits with $a$-cells, $X$ stabilisers with $(a-1)$-cells and $Z$ stabilisers with $(a+1)$-cells has parameters $[[n_a(C^{a,b}),\kappa^{a+b},\min(d^a,d^b)]]$.
Setting $a=b=1$ we recover the standard hypergraph product code construction, however in this work we will also consider the case $a=b=2$, obtaining 4D hypergraph product codes $HGP_{4D}(H)$.

\section{Belief propagation}\label{sec:bp}

The belief propagation (BP) algorithm (also known as \textit{sum-product algorithm}) has been shown to be effective at decoding classical LDPC codes~\cite{mackay1996near,mackay2003information}.
It has a running time that is linear in the block length  of the code.
In this section, we will review the use of BP for decoding a classical linear code with a binary $r\times n$ parity check matrix $H$.
We assume we have measured a syndrome $\mathbf{s}\in \mathbb{F}_2^r$, and the role of the decoder is to infer a likely error~$\mathbf{e}$, which must satisfy $H\mathbf{e}=\mathbf{s}$.
Although we consider classical linear codes in this section, BP can be directly applied to decode $Z$ (or~$X$) errors with CSS quantum codes by using the $X$ (or $Z$) parity check matrix in place of $H$, with the error $\mathbf{e}$ now corresponding to the $Z$ (or $X$) component of the error.
 
BP is a message passing algorithm, in which messages are passed along the edges of a Tanner graph. 
The Tanner graph is a graphical model describing a factorisation of the joint probability distribution of the error $\mathbf{e}$ and syndrome vector $\mathbf{s}$.
The factorised joint probability distribution $P(\mathbf{e},\mathbf{s})$ is
\begin{equation}\label{eq:joint_prob}
\begin{split}
P(\mathbf{e},\mathbf{s})&=P(\mathbf{e})\mathbbm{1}[\mathbf{s}=H\mathbf{e}]\\
&=\prod_{i}P(e_i)\prod_j\mathbbm{1}[s_j=\sum_{a\in\partial j}e_a]
\end{split}
\end{equation}
where $P(\mathbf{e})$ is the prior probability distribution over the noise vector (i.e.~$P(e_i)=p$ for a binary symmetric channel with a probability $p$ of error) and $\partial j\subseteq \{1,2,\ldots,n\}$ denotes the indices of the variables involved in the $j$\textsuperscript{th} parity check in $H$.
The Tanner graph $\mathcal{T}(H)$ of \Cref{eq:joint_prob} has a factor node $f_j$ for each parity check and a variable node $v_i$ corresponding to the random variable associated with each element $e_i$ of $\mathbf{e}$.
There is an edge between $f_j$ and $v_i$ in the factor graph if variable $i$ is involved in parity check $j$.

The problem BP approximately solves is
the bitwise decoding problem of finding the marginal posterior probabilities of each bit, $P(e_i=1|\mathbf{s})$. BP solves the bitwise decoding problem exactly on tree graphs, but is not exact on graphs that contain loops. 
While the factor graphs of LDPC codes do contain loops, BP can still perform well in practice provided the girth of the graph is sufficiently large.

For our implementation of BP, we represent each binary random variable $U$ using a log-likelihood ratio (LLR) defined as $L(U)=\log(P(U=0)/P(U=1))$.
BP involves repeating multiple iterations, where each iteration consists of a ``horizontal step'' and a ``vertical step''. 
In the horizontal step, which iterates over the rows of $H$, each parity check factor $f_j$ sends a message $Q_{f_j\rightarrow v_i}$ to its adjacent variable nodes $\{v_i:i\in\partial_j\}$:
\begin{equation}\label{eq:horiz_tanh}
Q_{f_j\rightarrow v_i}\coloneqq (-1)^{s_j}2\tanh^{-1}\left(\prod_{i^\prime\in\partial j\setminus i}\tanh \left(Q_{v_{i^\prime}\rightarrow f_j}/2\right)\right)
\end{equation}
where each $Q_{v_{i}\rightarrow f_j}$ is initialised before the first iteration to the LLR of the prior of variable $v_i$, which is $L(v_i)=\log((1-p)/p)$ for the binary symmetric channel with bit-flip probability $p$.

In the vertical step, which iterates over the columns of~$H$, each variable node $v_i$ sends a message $Q_{v_{i}\rightarrow f_j}$ to its adjacent factor nodes $\{f_j:j\in\partial_i\}$:
\begin{equation}\label{eq:vert_tanh}
Q_{v_i\rightarrow f_j}\coloneqq L(v_i)+\sum_{j^\prime\in\partial i \setminus j}Q_{f_{j^\prime}\rightarrow v_i}.
\end{equation}
We can also obtain an estimate $Q_{v_i}$ of the LLR of the posterior of each variable $v_i$ at each step:
\begin{equation}
Q_{v_i}\coloneqq L(v_i)+\sum_{j\in\partial i}Q_{f_{j}\rightarrow v_i}.
\end{equation}
These estimates of the posteriors of each variable are also called the \textit{soft decisions}.
From the soft decisions we then obtain \textit{hard decisions} $\mathbf{c}\in\mathbb{F}_2^n$, where $c_i\coloneqq 0$ if $Q_{v_i}>0$ and $c_i\coloneqq 1$ otherwise.
If $\mathbf{c}$ is a \textit{valid correction} satisfying $H\mathbf{c}=\mathbf{s}$ then BP stops and returns $\mathbf{c}$ as a correction.
If $H\mathbf{c}\neq \mathbf{s}$ then BP instead continues and runs the next iteration of the horizontal and vertical step.
If BP reaches a maximum number of iterations without finding a valid correction then it instead returns a heralded failure.
There are therefore two possible failure mechanisms from BP: it either returns a valid correction that removes the syndrome but leaves a residual logical error $\mathbf{e}+\mathbf{c}$, or it fails to find a valid solution, resulting in a heralded failure.
In our work we use a maximum number of iterations of 30 (this compares to 32 iterations used in Ref.~\cite{Panteleev2021degeneratequantum}). 
We do not observe significant improvements in decoding performance from further increasing the number of iterations.

The update rules in \Cref{eq:horiz_tanh} and \Cref{eq:vert_tanh}, which are called the ``tanh rule'', can result in numerical underflow issues when the log-likelihood ratios become large.
One way of avoiding this problem is to use the so-called \textit{Jacobian} approach instead, which is mathematically equivalent to the tanh rule but does not suffer from numerical underflow issues.
Another commonly used choice of updates is the min-sum update, which is an approximation to the tanh rule that has reduced complexity compared to either the tanh rule or the Jacobian method, at the cost of slightly reduced decoding performance.
We review these different variants of BP in \Cref{app:bp_update_rules}, and refer the reader to Ref.~\cite{chen2005reduced} for more details.

\section{Ordered Statistics Decoding}

Often, a large fraction of the errors that occur in BP are heralded failures, where the decoder runs out of iterations and fails to converge. When a heralded failure occurs in BP, the \textit{ordered statistics decoding} (OSD) technique can be used to post-process the soft decisions output by BP and find a good correction that is consistent with the syndrome~\cite{fossorier2001iterative}. This combination of belief propagation and orders statistics decoding (BP+OSD) has been shown to be a good general purpose decoder of quantum LDPC codes~\cite{Panteleev2021degeneratequantum,roffe2020decoding}. 

In the context of OSD, we will refer to the BP soft decisions as \textit{reliabilities} $\mathbf{r}\in \mathbb{R}^n$, which indicate the reliability of each bit of the hard decisions $\mathbf{c}$ output by BP.
In other words, $r_j>r_i$ implies that bit~$c_j$ is more reliable than~$c_i$.
Our definition of reliabilities $r_i\coloneqq Q_{v_i}$ is consistent with Ref.~\cite{Panteleev2021degeneratequantum}, except we use the LLR not the probability that the bit did not flip, but this is equivalent for OSD due to the monotonicity of the LLR function.
The first stage of OSD (called order-0 reprocessing or OSD-0) involves finding a set of $k$ bit positions (indices) that have high reliability and also correspond to $k$ linearly independent columns of the generator matrix $G$ of the code.
We call this set of $k$ indices, which we will shortly define more precisely, the \textit{most reliable information set} $T$.
Generally, an information set is any set of $k$ indices corresponding to $k$ linearly independent columns of the generator matrix $G$ of the code.

Given an observed syndrome $\mathbf{s}$ and an assignment of a bitstring in $\mathbb{F}_2^k$ to the bits of the correction $\mathbf{c}$ located in an information set, there is a unique choice of the remaining $n-k$ bits such that $H\mathbf{c}=\mathbf{s}$.
Therefore, in OSD-0, we only use the bits of the BP hard decisions that have support in an information set
that we consider most reliable (as determined by the soft decisions), and after fixing these $k$ bits, there is a unique choice of the remaining bits that is consistent with the syndrome.

In order to obtain $T$, we first sort the columns of $G$ in order of non-decreasing reliability from left to right, and call the resultant matrix $G'$.
The \textit{most reliable basis} $B_k$ for the column space of $G$ is the first $k=\rank(G)$ linearly independent columns of $G'$, from right to left.
The most reliable information set $T$ is the set of positions of the columns in $B_k$.
We can find $T$ using the parity check matrix $H$ of the code, rather than $G$.
To do so, first we sort the columns of $H$ in order of non-decreasing reliability, obtaining the matrix $H_1$.
We find the first $n-k$ linearly independent columns of $H_1$ (e.g.~by performing Gaussian elimination from left to right), and these columns constitute the least reliable basis $\Omega_{n-k}$ for the column space of $H$.
We denote the set of locations of the columns in~$\Omega_{n-k}$ by $S\subseteq\{1,2,\ldots, n\}$.
In Ref.~\cite{fossorier1998reliability} it was shown that~$T$ is the complement of~$S$, i.e.~$T=\{1,2,\ldots,n\}\setminus S$.

The encoding operator $\mathcal{E}(\mathbf{x}):
\mathbb{F}_2^k\rightarrow \mathbb{F}_2^n$ takes as input a choice of correction $\mathbf{x}$ for the $k$ bits in $T$, and returns the only correction operator $\mathbf{u}$ consistent with the syndrome $\mathbf{s}$ and choice of $\mathbf{x}$. More formally, the correction $\mathbf{u}=\mathcal{E}(\mathbf{x})$ is the unique choice of $\mathbf{u}$ that satisfies both $H\mathbf{u}=\mathbf{s}$ and $\mathbf{u}_{[T]}=\mathbf{x}$, where $\mathbf{u}_{[T]}$ is the subset of $\mathbf{u}$ containing only the elements located in $T$.

In OSD-0, we output the correction $\mathcal{E}(\mathbf{c}_{[T]})$ if BP fails to converge.
In order-$w$ reprocessing (OSD-$w$), we instead carry out a brute force search over a set $D$ of vectors $\mathbf{x}\in D$ with a small hamming distance from $\mathbf{c}_{[T]}$, and choose the correction $\mathbf{u}=\mathcal{E}(\mathbf{x})$ with smallest weight $|\mathbf{u}|$ among all $\mathbf{x}\in D$.
There are many possible choices of~$D$ for this restricted brute-force search, but we choose to search over all $\mathbf{x}$ that differ on any subset of the $w$ weakest bits of $\mathbf{c}_{[T]}$ (the $w$ weakest bits are those with the smallest reliabilities).
This version of OSD-$w$ was also used in Ref.~\cite{Panteleev2021degeneratequantum}.

\section{Single-shot quantum error correction}

The quantum circuit used to measure the stabilisers of quantum error correcting code will itself be faulty, and it is therefore necessary that any fault-tolerant error correction protocol should be able to handle both data qubit and syndrome measurement errors.
For a surface code of distance $d$, this can be achieved using $O(d)$ repetitions of the stabiliser measurements, using the entire syndrome history when decoding~\cite{dennis2002topological}.
However, \citeauthor{bombin2015single} showed that there exist classes of error correcting codes for which \textit{single-shot} error correction is possible~\cite{bombin2015single}.
With single-shot error correction, the decoder only uses information from a single round of noisy syndrome measurements.
While single-shot error correction in general leaves a small residual error after each round, it can be shown that this residual error is local and can remain bounded over multiple rounds of error correction, which is sufficient for fault-tolerant quantum computation.
Several families of quantum error correcting codes have been shown to exhibit single-shot quantum error correction, including the 3D gauge color code~\cite{bombin2015single,brown2016fault}, 3D and 4D toric and hypergraph product codes~\cite{kubica2019cellular, vasmer2021cellular,quintavalle2020single,breuckmann2016local,campbell2019theory,zeng2019higher}, 4D hyperbolic codes~\cite{breuckmann2020single}, and quantum expander codes~\cite{fawzi2020constant}.
A property of a code that has been shown to be sufficient for single-shot decoding under an adversarial noise model is \textit{confinement}~\cite{quintavalle2020single}.
A code has $(t,f)$-confinement if, for all errors $E$ with $|E|^{\mathrm{red}}\leq t$, the condition
\begin{equation}\label{eq:soundness}
f(|\sigma(E)|)\geq |E|^{\mathrm{red}}
\end{equation}
is satisfied, where $f:\mathbb{Z}\rightarrow\mathbb{R}$ is an increasing function with $f(0)=0$.
In other words, a code has confinement if, for errors with sufficiently small reduced weight, the minimum weight error consistent with the syndrome is bounded above by some function of the size of the syndrome.
\citeauthor{quintavalle2020single} proved that the 3D hypergraph product guarantees $X$-stabilisers with the confinement property~\cite{quintavalle2020single}.
However, an example of a code that is \textit{not} confined is the 2D toric code, since the syndrome of a single string of $Z$ errors (or $X$ errors) has weight two, regardless of the weight of the error.

Some single-shot codes have linear dependencies amongst the check operators, leading to syndromes becoming codewords of a classical linear code (called a \textit{metacode}) that can be used for syndrome repair.
These linear dependencies are not a requirement for a code to be single-shot (indeed quantum expander codes~\cite{fawzi2020constant} are single-shot and confined, but can have full rank check matrices), however a metacode can nevertheless be useful when decoding~\cite{brown2016fault,quintavalle2020single}.
We can construct a code that has syndromes encoded in a metacode using a chain complex with length at least three (to obtain a metacode for either $X$ or $Z$ syndromes), or length four if we would like a metacode for both $X$ and $Z$ syndromes.
Consider a chain complex with length $l$ at least four, and with qubits on the i-cells, where $1<i<l-1$. In addition to the $X$ check matrix $H_X\coloneqq\partial_i$ and $Z$ check matrix $H_Z\coloneqq\partial_{i+1}^T$, we additionally define the $X$ \textit{metacheck} matrix $M_X\coloneqq\partial_{i-1}$ and the $Z$ metacheck matrix $M_Z\coloneqq\partial_{i+2}^T$.
From \Cref{eq:bound2}, it must hold that $M_XH_X=0$, and hence any valid noiseless $X$ syndrome~$\mathbf{z}$ (in the image of $H_X$) must satisfy $M_X\mathbf{z}=0$. Similarly, the $Z$ metacheck matrix satisfies $M_ZH_Z=0$. 
We can therefore think of valid $X$ ($Z$) syndromes as codewords of a metacode with check matrix $M_X$ ($M_Z$).
Invalid syndromes in $\ker M_X$ belong to the $(i-1)$th homology group $\operatorname{H}_{i-1}\coloneqq \ker \partial_{i-1}/\im \partial_i=\ker M_X/\im H_X$. Invalid syndromes in $\ker M_Z$ belong to the $(i+1)$th cohomology group $\operatorname{H}^{i+1}$. 

\subsection{Two-stage decoding}

Given a noisy syndrome $\mathbf{z}^\prime$, one approach to finding a correction operator is to use a two-stage decoder. 
We will consider two-stage decoding of CSS codes, using the $X$ stabilisers and $X$ metacheck matrix to decode in the presence of $Z$ errors and $X$ stabiliser measurement errors.
Exactly the same method can be used to decode $X$ errors with $Z$ stabiliser measurements.

In the first stage of a two-stage decoder, the metacheck matrix $M_X$ is first used to find the \textit{metasyndrome} $\mathbf{s}=M_X\mathbf{z}^\prime$. 
A decoder $f_{d}^1:\mathbb{F}_2^{n_{i-2}}\rightarrow \mathbb{F}_2^{n_{i-1}}$ is then used to find a probable noiseless syndrome $\mathbf{z}$ given $\mathbf{s}$.
In the second stage another decoder $f_d^2:\mathbb{F}_2^{n_{i-1}}\rightarrow \mathbb{F}_2^{n_i}$ is used to find a probable noise vector $\mathbf{n}$ given the corrected syndrome $\mathbf{z}$, such that $H_X\mathbf{n}=\mathbf{z}$.

There are two ways that such a correction can fail. Either the corrected syndrome $\mathbf{z}=f_d^1(\mathbf{s})$ is invalid, belonging to $H_{i-1}$, or the correction operator $\mathbf{n}=f_d^2(\mathbf{z})$ belongs to $H_i$. In Ref.~\cite{quintavalle2020single}, the authors remove the first failure mechanism, which they call a \textit{metacode failure}, by using a modified metacheck matrix
\begin{equation}
M^{\prime}=\left(\begin{array}{c}
M_X \\
L_{M}
\end{array}\right)
\end{equation}
where the rows of $L_{M}$ generate the $(i-1)$th cohomology group $\operatorname{H}^{i-1}$. Note that if $L_M$ is not sparse (as is typically the case for topological codes, for example), then decoding using $M^\prime$ may be harder for some decoders (such as belief propagation). As a result, in Ref.~\cite{quintavalle2020single}, the authors only use $M^\prime$ if the first attempt at decoding $\mathbf{s}=M\mathbf{z}^\prime$ yields an invalid syndrome.
However, as noted in Ref.~\cite{quintavalle2020single}, the threshold of two-stage decoders can be limited by that of the metacode, leading to performance that is far from optimal.

\subsection{Single-stage decoding}

\begin{figure}
\centering
\includegraphics{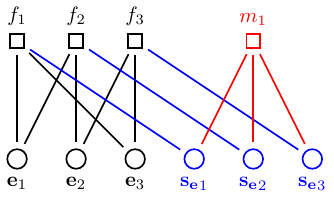}
\caption{The Tanner graph of a repetition code with noisy syndrome measurements. Also shown is the metacheck $m_1$, which is just the linear combination $m_1=f_1+f_2+f_3$ of the other three parity checks.}\label{fig:syndrome_variable_factor_graph}
\end{figure}

Suppose a $Z$-type data qubit error $\mathbf{e}$ and syndrome measurement error $\mathbf{s_e}$ have occurred, giving the observed syndrome $\mathbf{s}=H_X\mathbf{e}+\mathbf{s_e}$.
Rather than decoding in two stages, a natural decoding strategy we might instead consider is to find the most probable data qubit and syndrome measurement error consistent with the observed syndrome, given the error model for data qubit errors and syndrome measurement errors.
This decoding problem can be expressed as finding the minimum weight error
\begin{equation}
\mathbf{e}'\coloneqq\left(\begin{array}{c}
\mathbf{e} \\
\mathbf{s_e}
\end{array}\right)
\end{equation}
satisfying $H^{\prime}\mathbf{e}'=\mathbf{s}$ where $H^{\prime}$ is now the modified parity check matrix defined as
\begin{equation}\label{eq:augmented_Hx}
H^{\prime}=\left(\begin{array}{cc}
H_X & I_r
\end{array}\right).
\end{equation}
The problem of minimum weight decoding of general linear codes is known to be NP-complete~\cite{berlekamp1978inherent}, however we can still apply alternative decoding algorithms to the modified parity check matrix $H^{\prime}$. 
Before discussing the choice of decoder in more detail, let us first consider the structure of Tanner graph $\mathcal{T}(H^\prime)$ of $H^{\prime}$.
The graph $\mathcal{T}(H^\prime)$ is obtained from $\mathcal{T}(H_X)$ by adding a variable node $v^m_i$ for each check node $f_i$, as well as adding the edge $(v^m_i,f_i)$.
These new nodes and edges are highlighted in blue in the repetition code example in \Cref{fig:syndrome_variable_factor_graph}.
Each of these additional variable nodes corresponds to a bit flip error mechanism in the syndrome measurement outcome of the check node it is connected to.

This inclusion of a variable node for each parity check in the Tanner graph to handle measurement errors in the context of decoding with belief propagation was introduced in Ref.~\cite{li2020numerical} to decode Bravyi-Bacon-Shor and subsystem hypergraph product codes, and has also been used to decode hypergraph product codes~\cite{grospellier2020combining, kuo2021decoding} and 4D hyperbolic codes~\cite{breuckmann2020single}.
In our work, we make two additional modifications.
Firstly, as we show in \Cref{sec:numerics}, belief propagation on its own is not sufficient for single-shot decoding of 3D and 4D toric codes, owing to the presence of degenerate error configurations leading to the split-beliefs problem.
Therefore, we use OSD post-processing to fix this degeneracy problem~\cite{Panteleev2021degeneratequantum}.
Secondly, since we have metachecks for our codes, we include these explicitly on the Tanner graph.
These metachecks act solely on the newly added variable nodes corresponding to measurement errors.
An example of a metacheck is shown in red in \Cref{fig:syndrome_variable_factor_graph}.
With the metachecks added, our new Tanner graph $\mathcal{T}(H^M)$ corresponds to the check matrix
\begin{equation}\label{eq:check_matrix_with_metachecks}
H^M=\left(\begin{array}{cc}
H_X & I_r \\
0 & M
\end{array}\right)
\end{equation}
where here $M$ is the metacheck matrix.
Note that the rows corresponding to the metachecks (the bottom half of $H^M$) are linear combinations of the rows in the top half, since $M(H_X, I_r)=(0, M)$.
As such, the metachecks are already implicitly present in $\mathcal{T}(H^\prime)$ as linear combinations of other checks, however we find that adding them explicitly to the Tanner graph and using the syndrome $(\mathbf{s}, M\mathbf{s})^T$ improves the decoding performance of BP+OSD.
We note that both \Cref{eq:check_matrix_with_metachecks} and \Cref{eq:augmented_Hx} can be interpreted as the $X$ check matrix of a data-syndrome code~\cite{ashikhmin2014robust,kuo2021decoding}.

Note that, since the BP+OSD decoder finds a solution~$\mathbf{e}'$ satisfying $H^\prime \mathbf{e}'=\mathbf{s}$, we are guaranteed to have $\mathbf{s}+\mathbf{s_e}\in\im H_X$, and so metacode failures are not possible with this single stage decoding.
This is in contrast to two-stage decoding, where the $L_M$ matrix must be included in the modified metacheck matrix $M'$ to avoid metacode failures.
However, since $L_M$ is typically not sparse, this method can degrade the performance of decoders such as belief propagation.

\section{Numerical simulations}\label{sec:numerics}

We simulated the performance of single-stage and two-stage decoders under a phenomenological noise model, in which we perform $N$ rounds of noisy syndrome measurement, followed by a final round of noiseless syndrome measurements. 
In each round of measurements, each data qubit suffers a $Z$ (or $X$) error with probability $p$. In the rounds of noisy syndrome measurements, the stabiliser is also measured incorrectly with probability $p$.
After each round of stabiliser measurements, we use our single-shot decoder to apply a correction operator.
The final round of perfect stabiliser measurements simulates logical measurement by measuring all data qubits destructively in either the $X$ or $Z$ basis.
When each data qubit is measured destructively, measurement errors have the same effect as data qubit errors, and stabilisers can be determined without error from classical post-processing.
For the majority of our numerical simulations we use our own implementation of BP+OSD for simulations, which is consistent with Ref.~\cite{Panteleev2021degeneratequantum} using the exhaustive search strategy for post-processing with $w=10$.
The only results which use a different implementation are those that use the combination sweep strategy in \Cref{sec:limitations_of_bp_osd}, for which we used the LDPC Python package~\cite{Roffe_LDPC_Python_tools_2022}.

\subsection{Decoder comparison}\label{sec:decoder_comparison}

\begin{figure}
    \centering
    \includegraphics[width=0.8\columnwidth]{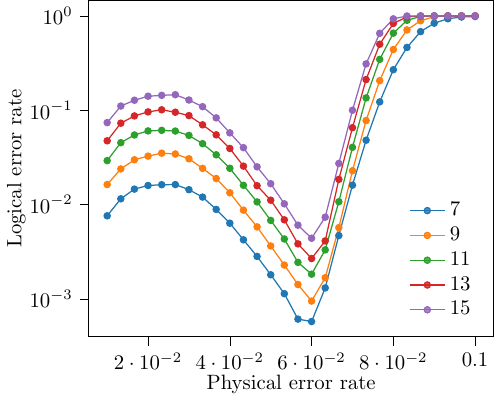}
    \caption{Decoding performance of the single-stage belief propagation decoder (without metachecks or OSD) for the 3D toric code, using 8 rounds of noisy syndrome measurement.}
    \label{fig:bp_only_single_shot}
\end{figure}

In \Cref{fig:bp_only_single_shot} we show the performance of the single stage BP decoder (without OSD or metachecks) for the 3D toric code, with 8 rounds of noisy syndrome measurements.
Clearly BP does not have a threshold and, unusually, there is a regime where the logical error rate actually \textit{increases} as the physical error rate decreases (with a peak in logical error rate at physical error rates of around 2\% to 3\%).

\begin{figure}
    \centering
    \includegraphics[width=1.0\columnwidth]{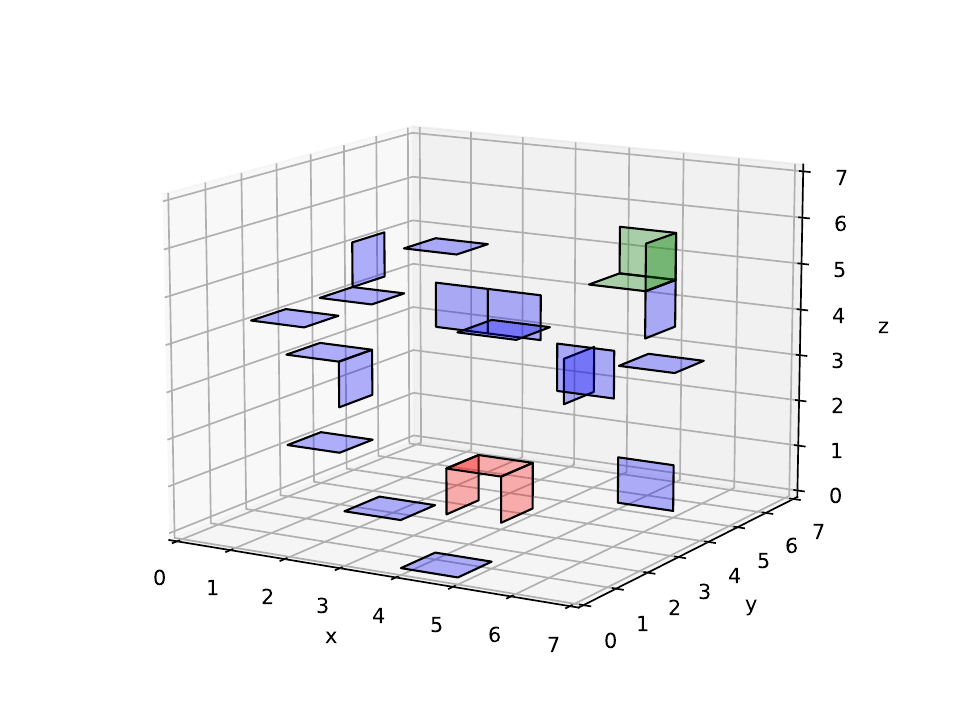}
    \caption{Example of a $Z$ error in the $L=7$, 3D toric code, with qubits identified with faces. 
    A single round of $Z$ noise on data qubits was simulated with perfect syndrome measurements. Highlighted in red is an isolated ``half-cube'' error which leads to split beliefs when decoded with BP. Highlighted in green is another half-cube error which is decoded successfully by BP, since the adjacent (blue) error breaks the degeneracy. }
    \label{fig:bp_error_visualised}
\end{figure}

In order to understand this behaviour better, we visualised typical errors that lead to BP failing when decoding the 3D toric code for one round of perfect syndrome measurements (since we observed similar behaviour with this simpler error model).
An example of such an error is shown in \Cref{fig:bp_error_visualised}.
We find that BP can fail to converge when a ``half-cube'' failure mechanism is present: that is, when a $Z$ error occurs on three sides of a cube.
If we denote such an error as~$E$, and then denote the $Z$ stabiliser associated with the same cube as $S$, we note that~$E$ and~$SE$ both have the same syndrome and occur with the same error probability.
This can lead to the problem of ``split beliefs'' resulting from degenerate errors, where the marginal posteriors output by BP are split between equally probable errors, leading to hard decisions that are inconsistent with the syndrome~\cite{poulin2008iterative}.
However, we observe that other errors in close proximity to a half-cube can break the symmetry in the decoding problem, allowing BP to converge.
This is illustrated in \Cref{fig:bp_error_visualised} where the half-cube highlighted in green is decoded correctly by BP (and is adjacent to another error), whereas the half-cube highlighted in red is isolated and cannot be decoded by BP alone owing to the problem of split beliefs.

\begin{figure}
    \centering
    \includegraphics[width=0.9\columnwidth]{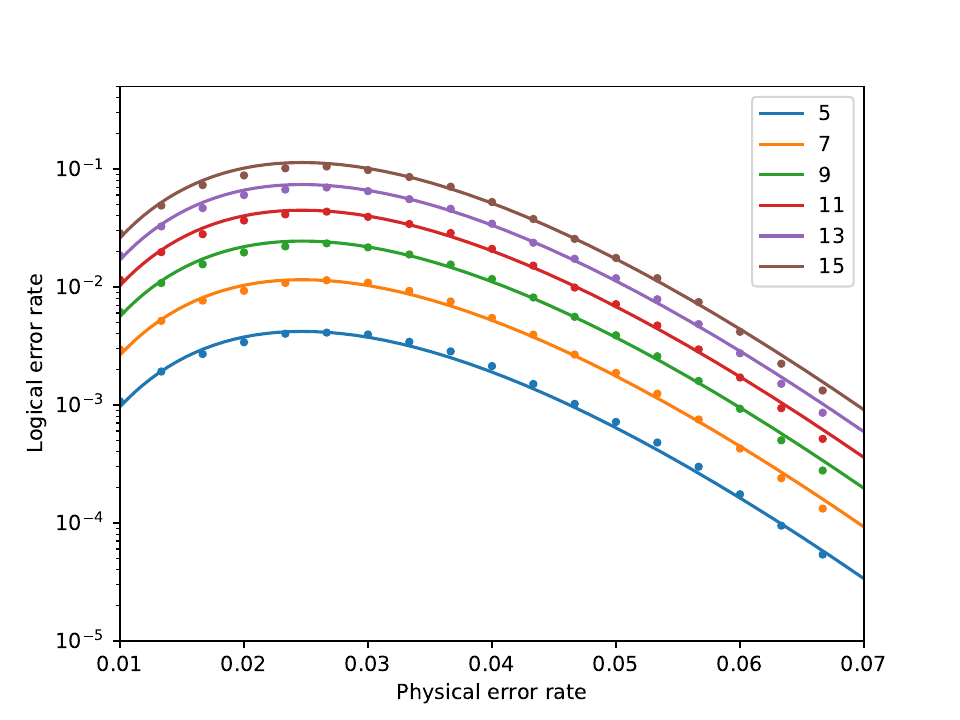}
    \caption{Individual markers (circles) show the performance of the 3D toric code using the belief-propagation decoder with perfect syndrome measurements. The solid line shows $0.42\times$ the probability that a ``half-cube'' error occurs surrounded by a region of 245 qubits of which at most 4 suffer an error (setting $\alpha=0.42$, $r=245$ and $q_{\max}=4$ in \Cref{eq:isolated_half_cube_probability}). The legend gives the lattice size $L$.}
    \label{fig:model_of_bp_only_failures}
\end{figure}

We model the probability of this failure mechanism by the probability that an isolated half-cube error occurs. The probability that a half-cube error occurs on a specific cube of the lattice is ${6 \choose 3}p^3(1-p)^3$, and the probability that $q$ errors occur in a neighbouring region of $r$ qubits is ${r \choose q}p^q(1-p)^{(r-q)}$. We assume that only some fraction $\alpha$ of these error configurations lead to BP failing to converge.
Further, we assume that the error with support in the rest of the lattice (outside the neighbouring region of a half-cube) has no impact on BP decoding success.
For an $L\times L\times L$ 3D toric code, there are $L^3$ locations where these isolated half-cube errors can be centred.
Assuming that up to $q_{\max}$ errors can occur in the neighbourhood surrounding a half-cube, this leads to an estimated BP logical failure probability of
\begin{equation}\label{eq:isolated_half_cube_probability}
    p_{\mathrm{log}}^{\mathrm{est}} = 20\alpha L^3 \sum_{q=0}^{q_{\max}} {r \choose q}p^{(q+3)}(1-p)^{(r-q+3)}
\end{equation}
where $q_{\max}$, $r$ and $\alpha$ can be determined through a fit to the simulated data.
We find a good fit with $\alpha=0.42$, $r=245$ and $q_{\max}=4$ (see \Cref{fig:model_of_bp_only_failures}), which supports the conclusion that isolated half-cube errors are a dominant failure mechanism for BP in the 3D toric code.
We note that the value $r=245$ is close to the number of faces (240) in a $4\times 4\times 4$ cube in the lattice.
Our \Cref{eq:isolated_half_cube_probability} overcounts some error configurations, such as where more than one half-cube error occurs in the lattice.
However, we still expect it to be an reasonable estimate of the probability of this class of errors occurring to leading order.

\begin{figure}
    \centering
    \includegraphics[width=0.8\columnwidth]{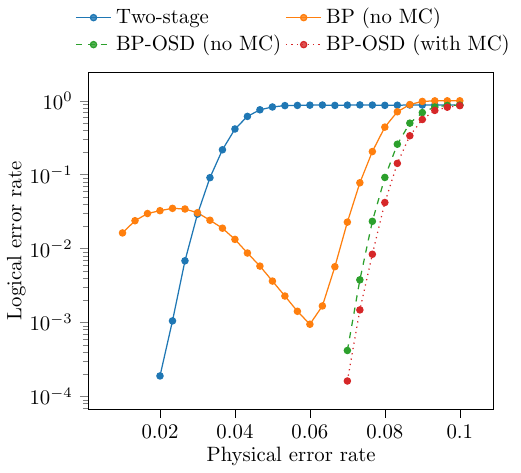}
    \caption{Comparison of decoders for the $L=9$ 3D toric code with 8 rounds of noisy syndrome measurements. We compare a two-stage decoder (blue) with three single-stage decoding strategies using either BP alone or BP-OSD. We compare the performance of single-stage decoding with metachecks either included explicitly in the Tanner graph (``with MC'') or omitted (``no MC'').}
    \label{fig:compare_decoders}
\end{figure}

We compare different decoding strategies for an $L=9$ 3D toric code in \Cref{fig:compare_decoders}.
While single-stage BP decoding does outperform the two-stage decoder (using BP-OSD) at higher physical error rates (e.g.~for error rates of around 3\% to 8\%), it performs worse for error rates below 3\% owing to the dominant half-cube errors.
Fortunately, half-cube errors are heralded errors, where we know that BP has failed to converge (BP outputs a correction inconsistent with the syndrome).
As a result, these half-cube errors can be fixed using OSD post-processing, which uses the BP posterior probabilities to guide the choice of a correction consistent with the syndrome.
As shown in \Cref{fig:compare_decoders} we observe that single-stage decoding with BP-OSD does indeed significantly outperform single-stage decoding with BP alone, and does not suffer from the increase in logical error rate at low physical error rates due to half-cube errors.
We also achieve a slight improvement in performance by adding metachecks explicitly into the Tanner graph used by BP to decode.

\subsection{Sustainable thresholds}

After each round of noisy stabiliser measurements, we will not in general be able to find a correction operator that returns the state to the codespace.
However, we can still protect quantum information for an arbitrarily long time period provided that the error from decoding the noisy syndrome in each round is sufficiently small to be corrected in subsequent rounds.
This motivates the definition of the \textit{sustainable threshold}.
The sustainable threshold of a single-shot code family is the physical error rate $p_{sus}$ below which quantum information can be stored indefinitely by increasing the distance of the code, and is defined to be~\cite{brown2016fault}
\begin{equation}
p_{sus}=\lim_{N \to \infty}p_{th}(N)
\end{equation}
where $p_{th}(N)$ is the threshold for $N$ rounds of stabiliser measurements. 
In Refs.~\cite{brown2016fault,quintavalle2020single}, the authors determine $p_{sus}$ by fitting to the ansatz
\begin{equation}\label{eq:p_sus_exp_ansatz}
p_{th}=p_{sus}[1+(p_0/p_{sus}-1)e^{-\gamma N}],
\end{equation}
however we do not find this ansatz to be a good fit for the code families and decoders we consider in this work.
We found that an ansatz of the form $p_{th}(N)=p_{sus}(1 + (p_0/p_{sus} - 1)/(N+1))$ approximated our data much better, but still not closely enough.
Therefore, we instead determine $p_{sus}$ by estimating $p_{th}(N)$ numerically for increasing values of $N$ (doubling $N$ each time we increase it), until increasing $N$ no longer decreases $p_{th}(N)$ to within the statistical significance of our threshold estimates.

\begin{figure}
\includegraphics[width=0.8\columnwidth]{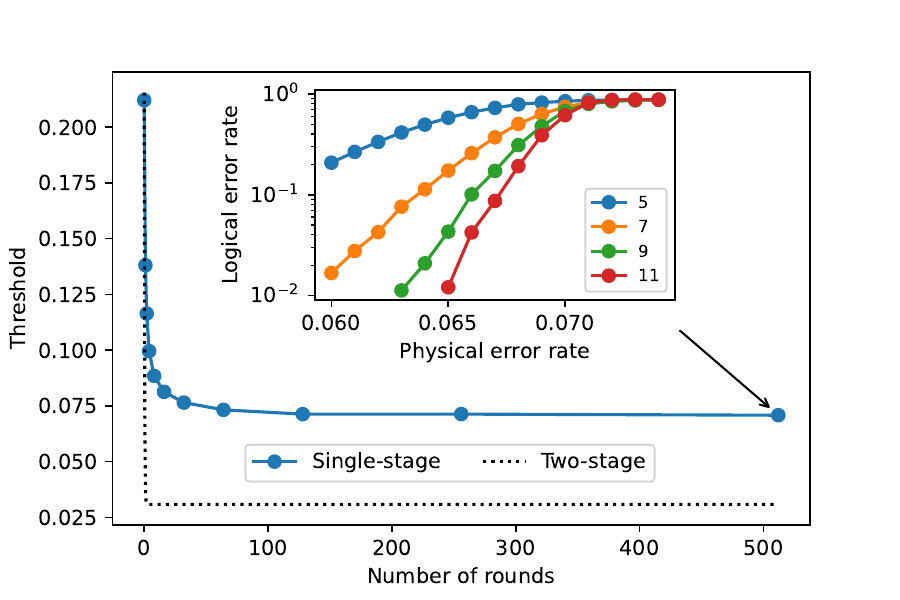}
\caption{Single-shot sustainable $Z$-threshold estimate for the 3D toric code using the single-stage BP+OSD decoder, compared to results for the two-stage BP+OSD+MWPM decoder from Ref.~\cite{quintavalle2020single}. The threshold is plotted as a function of the number of noisy measurement rounds before the final round of perfect measurements. Inset is the threshold plot for 512 noisy rounds of syndrome measuremen, with the lattice size $L$ given in the legend.}
\label{fig:single_stage_vs_two_stage_3d}
\end{figure}

\begin{figure}
\includegraphics[width=0.8\columnwidth]{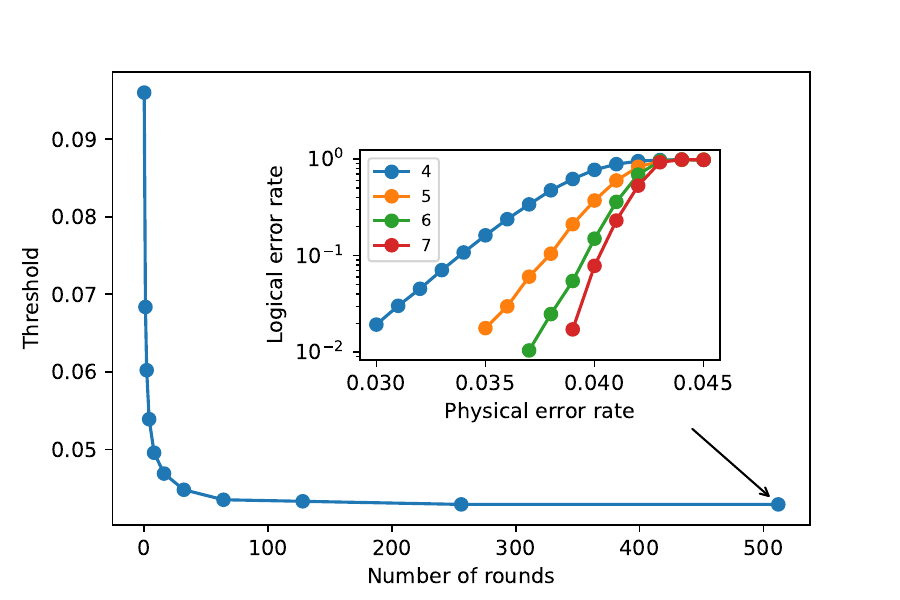}
\caption{Single-shot sustainable threshold estimate for the 4D toric code using the single-stage BP+OSD decoder. The threshold is plotted as a function of the number of noisy measurement rounds before the final round of perfect measurements. We estimate a sustainable threshold of 4.3\%. Inset is the threshold plot for 512 noisy rounds of syndrome measuremen, with the lattice size $L$ given in the legend of the subplot.}
\label{fig:single_shot_2d_4d}
\end{figure}

For the 3D toric code, with qubits on 2-cells, our results are consistent with a single-shot sustainable threshold for $Z$ errors of at least $7.1\%$ using the single-stage BP+OSD decoder, as shown in \Cref{fig:single_stage_vs_two_stage_3d}.
Since the logical failure rate is saturated at the threshold, it may be that the crossing occurs at a higher physical error rate (our estimate of 7.1\% is a lower bound).
This significantly outperforms the threshold of $3.08\%$ found for two-stage decoding with BP+OSD+MWPM in Ref.~\cite{quintavalle2020single}. 
In \Cref{fig:single_shot_2d_4d} we plot $p_{th}(N)$ for the 4D toric code (with qubits on 2-cells) using the single-stage BP+OSD decoder, and using lattice sizes $L\in\{4, 5, 6,7\}$. 
These results are consistent with a sustainable threshold of at least $p_{sus}=4.3\%$, and again we find that the logical failure rate is saturated at threshold. 
Since the 4D toric code is self-dual, the 4.3\% sustainable threshold we find is both the $Z$ and $X$ threshold. 
To the best of our knowledge, this is the highest single-shot threshold found to date for any code family, and even exceeds the $2.93(2)\%$ threshold of the 2D toric code for the same noise model but using $L$ rounds of stabiliser measurements~\cite{wang2003confinement}.

\subsection{Overhead reduction}

\begin{figure}
\includegraphics[width=0.8\columnwidth]{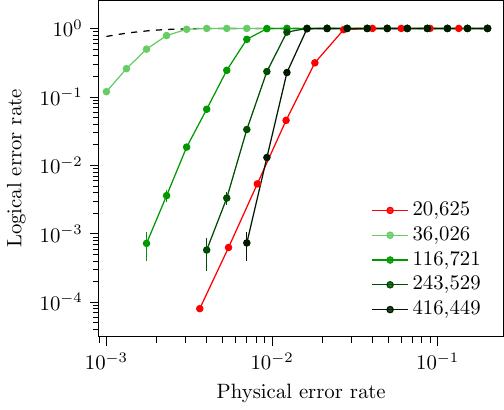}
\caption{Performance of a [[20625,1441,9]] 4D hypergraph product code constructed from a [10,6,3] linear code decoded using BP (red), compared to 1441 copies of $L\in \{5,9,13,17\}$ rotated surface codes (shades of green) decoded with minimum-weight perfect matching. All codes are simulated under a phenomenological noise model for 32 rounds of noisy syndrome measurement, and one final round of perfect syndrome measurements. The number of physical qubits used are given in the legend, and the dashed line shows the error rate for 1441 unencoded physical qubits.}
\label{fig:hgp_4d_vs_surface}
\end{figure}

We also investigated the performance of 4D hypergraph product codes with improved code parameters relative to the surface code. In \Cref{fig:hgp_4d_vs_surface} we show the performance of a [[20625,1441,9]] 4D hypergraph product code, constructed from the fourfold tensor product of a [10,6,3] linear code.
The $[10,6,3]$ was constructed using the method given in~\cite{nealldpc} using check matrices of dimensions $(5,10)$ and a column weight of 2 (we constructed 100 check matrices this way and chose the code that performed best when decoded with BP).
We computed the distance using the result in Ref.~\cite{zeng2019higher}.
In this case, we decoded using BP (no OSD), and included the metachecks explicitly in the Tanner graph.
The decoding performance without OSD is better than for other block codes we have tried, since there are fewer degenerate errors, owing to the higher weight stabilisers compared to the 4D toric code.
The [[20625,1441,9]] code has stabiliser weight 8 and 10, compared to 6 for the 4D toric code.
For a phenomenological noise model, using 32 rounds of noisy syndrome measurement, we find that the [[20625,1441,9]] code outperforms rotated surface codes of a similar size, even though the full syndrome history is used by MWPM for the surface codes when decoding.
For a physical error rate of around 1\%, the [[20625,1441,9]] 4D hypergraph product code requires between $10\times$ and $20\times$ fewer physical qubits than the surface code for the same logical error rate and number of logical qubits.
We note that the logical error rate of the [[20625,1441,9]] 4D hypergraph product code has a similar slope in \Cref{fig:hgp_4d_vs_surface} to the 1441 copies of distance 9 surface codes (requiring 116,721 qubits), which is consistent with the [[20625,1441,9]] being decoded up to the full code distance.
This also suggests that the reduction in qubit overhead might reduce to around $116721/20625\approx5.7\times$ in the limit of low physical error rates and low (e.g.~$10^{-12}$) target logical error rates.

\begin{figure}
\includegraphics[width=0.8\columnwidth]{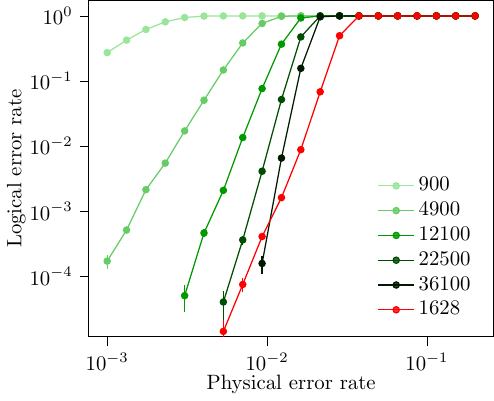}
\caption{Performance of a [[1628,100]] balanced product code, constructed from the product of expander codes derived from a Cayley graph of the cyclic group~$C_{22}$ with a [14,9,4] local code. The balanced product code (red) is compared to rotated surface codes (green). In order to keep the number of logical qubits fixed in the comparison, we take 100 copies of the rotated surface codes (we plot $1-(1-p_{log})^{100}$). All codes are simulated for 32 rounds of noisy measurements and for the balanced product code we use a single-stage single-shot BP+OSD decoder (with $w=10$).}
\label{fig:balanced_product_vs_rotated_surface_codes}
\end{figure}

We also investigate the performance of balanced product codes, which have been shown to have asymptotically better parameter scaling than hypergraph product codes~\cite{breuckmann2021balanced}.
In \Cref{fig:balanced_product_vs_rotated_surface_codes} we compare the performance of a balanced product code with rotated surface codes. 
The balanced product code has parameters [[1628,100]], and is constructed from the balanced product of an expander code and its dual. 
The Tanner code is derived from the Cayley graph of the cyclic group $C_{22}$ with degree 14 with a [14,9,4] block code (from~\cite{Grassl:codetables}) as the local code.
To choose the generating set $S$ of the Cayley graph of $C_{22}$, we randomly sampled 200 generating sets and chose the set that produced a Cayley graph with the minimum $\lambda_2$.
Here $\lambda_2$ is the second-largest eigenvalue of the adjacency matrix of the graph and, due to the Cheeger inequalities, a smaller $\lambda_2$ implies that the graph has better expansion (see Section III of \cite{breuckmann2021balanced} for a review of how $\lambda_2$ relates to the properties of expander codes).
The generating set we use has $\lambda_2=1.92$.
The $X$ and $Z$ stabilisers both have check weights at least 8 and at most 13.
We find a $10\times$ to $20\times$ reduction in qubit error rate at a physical error rate of around 1\%.
This is a similar reduction in overhead as achieved by the 4D hypergraph product code in \Cref{fig:hgp_4d_vs_surface}, however we note that the balanced product code achieves this overhead reduction with a much smaller block length.
Our results in \Cref{fig:balanced_product_vs_rotated_surface_codes} provide evidence that balanced product codes can be single-shot, and we provide additional numerical results that also support this conclusion in \Cref{sec:balanced_product_vary_noisy_rounds}.

\subsection{Limitations of BP+OSD}\label{sec:limitations_of_bp_osd}

\begin{figure}
\includegraphics[width=0.8\columnwidth]{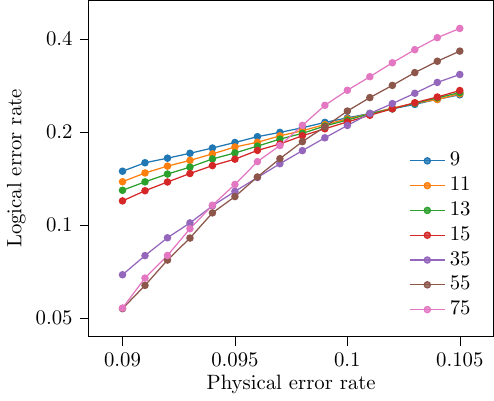}
\caption{Performance of BP+OSD for decoding the 2D toric code. We used the Ref.~\cite{Roffe_LDPC_Python_tools_2022} implementation of BP+OSD with the combination sweep strategy for OSD (setting $\lambda=60$)~\cite{roffe2020decoding}, and product sum updates for BP with a maximum number of iterations equal to the code block length. The lattice size $L$ is given in the legend.}
\label{fig:bp_osd_2d_toric}
\end{figure}

\begin{figure}
\includegraphics[width=0.8\columnwidth]{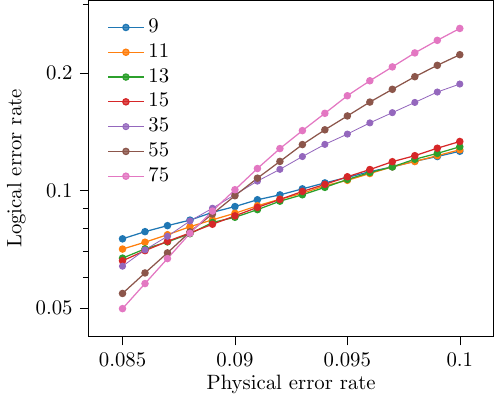}
\caption{Logical $Z$ error rate of the triangular color code decoded using BP+OSD for a code capacity noise model, with the same implementation and settings of BP+OSD as used for \Cref{fig:bp_osd_2d_toric}. The lattice size is given in the legend.}
\label{fig:bp_osd_2d_color_code}
\end{figure}

BP+OSD has already been shown to have remarkably good decoding performance for a wide range of quantum LDPC codes in the absense of syndrome measurement errors~\cite{Panteleev2021degeneratequantum,roffe2020decoding}, and in this work we have shown how it can be used to obtain good single-shot decoding performance for higher dimensional hypergraph product codes.
However, BP+OSD has limitations that present a challenge for its use as a general-purpose decoder for quantum LDPC codes.
Firstly, BP+OSD has a running time time that is at least cubic in the size of the parity check matrix.
This is significantly more computationally expensive than the union-find decoder for the surface code~\cite{delfosse2017almost}, and as a result real-time hardware implementations of BP+OSD may be difficult to achieve~\cite{valls2021syndrome}.
Secondly, we have observed that the performance of BP+OSD can degrade for 2D toric and color codes for large system sizes, as shown in \Cref{fig:bp_osd_2d_toric} and \Cref{fig:bp_osd_2d_color_code}.
Although BP+OSD is unlikely to be used for the 2D toric and color codes, as these codes already have efficient decoders, it is nevertheless useful to consider them for benchmarking purposes.
If only small system sizes are considered ($L\leq 15$), the results of \Cref{fig:bp_osd_2d_toric} are consistent with a BP+OSD toric code threshold approaching that of minimum-weight perfect matching, as shown in Ref.~\cite{roffe2020decoding}.
However, for larger system sizes we find that the crossing point recedes significantly, which suggests either that the threshold is much lower than results for small system sizes might suggest, or that there may not be a threshold.
We observe similar behaviour for the 2D triangular color code, as shown in \Cref{fig:bp_osd_2d_color_code}.
For both \Cref{fig:bp_osd_2d_toric} and \Cref{fig:bp_osd_2d_color_code} we used the variant of OSD implemented in Ref.~\cite{Roffe_LDPC_Python_tools_2022} with the ``combination sweep'' post-processing strategy, setting $\lambda=60$, which involves a larger brute-force search than OSD-10 and a relative reduction in the logical error rate. 
In order to ensure that the decoding performance was not limited by the number of BP iterations, we used a maximum number of iterations equal to the block length of the code.
However, we did not observe a significant improvement from letting the number of iterations grow with block length, and in \Cref{sec:bp_locality} we argue that there may be limited benefit to increasing the number of iterations beyond some constant for Tanner graphs with a small constant girth, which justifies our choice of 30 iterations used in the remainder of this work.

One method of recovering a threshold with BP+OSD using the exhaustive search strategy is by allowing $w$ to grow with the block length, although this can lead to an exponential running time (since the brute-force search in OSD-$w$ has complexity $O(2^w)$). 
To understand why this approach recovers a threshold, consider an $X$ check matrix $H$ which we will decode with BP+OSD.
If we use OSD-$w$ post-processing, setting $w$ to be the nullity of $H$, then BP+OSD decoding with OSD-$w$ is equivalent to finding the minimum weight $Z$-type error, which is the problem that the minimum-weight perfect matching (MWPM) decoder already solves efficiently for the 2D toric code, with a threshold of 10.3\%~\cite{wang2003confinement}.
We would therefore ideally like to find a variant of OSD post-processing that has polynomial running time as well as a threshold both for the 2D toric code and more general families of quantum LDPC codes.
Unfortunately, the greedy search methods for OSD in the literature do not appear to meet this criteria, however one possible solution might be to use the BP posteriors to guide cluster growth in the generalised Union-Find decoder for quantum LDPC codes introduced in Ref.~\cite{delfosse2022toward}.
A new post-processing method called stabiliser-inactivation has recently been shown to improve on OSD for some quantum LDPC codes~\cite{crest2022stabilizer}, although its performance for the toric code has not yet been studied.

The 2D toric and color codes are amongst the most challenging codes to decode for BP+OSD, since they have many low-weight stabilisers of even weight that lead to low-weight degenerate error configurations. For example, any weight-2 $X$-type error $P$ in the support of a single site $X$ stabiliser $S$ in the toric code will have the same error probability and syndrome as the error $PS$, and will likely lead to BP failing to converge owing to the split-belief problem. 
For the 3D and 4D toric code, typical problematic error configurations that lead to split beliefs occur with lower probability.
This is a result of stabilisers having a higher weight than the 2D toric dcode, as well as a larger boundary in the Tanner graph than both the 2D toric and color codes, which increases the probability that nearby errors will break the degeneracy, as discussed in \Cref{sec:decoder_comparison}.
This makes the decoding problem solved by the OSD post-processing more challenging for the 2D toric and color codes than for the 3D or 4D toric codes. 
Indeed, we do not observe the crossing recede for the 3D and 4D toric codes, even though the distances and system sizes we consider are already quite large.
For example, the $L=7$ 4D toric code in \Cref{fig:single_shot_2d_4d} has 14,406 data qubits, and distance 49.
This compares to the only 6,050 data qubits in the $L=65$ 2D surface code in \Cref{fig:bp_osd_2d_toric}, for which decoding performance has degraded.
Will the 3D and 4D toric codes suffer the same problem of degraded performance with BP+OSD but just for larger lattice sizes than for the 2D toric code?
Testing this numerically is challenging, given the $O(n^3)$ running time of OSD.
Regardless, our numerical results demonstrate that single-stage single-shot decoding with BP+OSD for the 3D and 4D toric code already has good performance for very large codes (larger than would likely be used in practice), and so its performance in the asymptotic limit may not be so important in practice.

\section{Conclusion}

In this work, we have studied the problem of decoding higher dimensional hypergraph product codes in the single-shot regime.
A common decoding strategy for these higher dimensional codes has been to use a two-stage decoder, which first attempts to repair the syndrome, before using the corrected syndrome to estimate a data qubit correction.
However, two-stage decoders introduce additional failure mechanisms (metacheck failures), and can be bottlenecked by the syndrome repair stage.
Here, we demonstrate the improved performance that can be attained by decoding in a single stage.
Using a single-stage BP+OSD decoder, our results are consistent with a sustainable threshold of around 7.1\% for $Z$ errors with the 3D toric code with a phenomenological noise model, a significant improvement on the threshold of 2.90\% observed using the two-stage decoder of Ref.~\cite{quintavalle2020single}.
Since the optimal threshold of the syndrome repair step is 3.3\% for the 3D toric code~\cite{ohno2004phase}, our work confirms that this first stage was indeed limiting decoding performance~\cite{quintavalle2020single}.
For the 4D toric code, which is single-shot for both $X$ and $Z$ errors, our results are consistent with a threshold of around 4.3\% for a phenomenological noise model, which far exceeds the highest previously observed single-shot threshold of 1.59\%~\cite{breuckmann2016local} for this code, using the Hastings decoder~\cite{hastings2013decoding}.
Furthermore, this even exceeds the 2.93\% threshold of the 2D toric code for an equivalent noise model, but instead using $L$ repeated rounds of syndrome extraction.

Although we show that BP+OSD is effective when used for single-shot decoding of higher dimensional codes for the system sizes we considered, we also demonstrate that the performance of BP+OSD significantly degrades for large lattice sizes for the 2D toric code.
The 2D toric code is particularly challenging to decode with BP since it has many low weight stabilisers, leading to a high probability that there will be degenerate solutions for any given syndrome, which in turn causes the problem of ``split beliefs'' for BP~\cite{poulin2008iterative}.
We do not observe this degradation of performance for the higher dimensional codes we studied, despite simulating larger system sizes for these codes than for the 2D toric code.
However, we cannot rule out that a similar problem will not arise for even larger system sizes with these higher dimensional codes, and it is challenging to test this regime numerically given the high computational complexity of OSD, which is cubic in the number of qubits.
Our results therefore motivate further work to prove the existence of a threshold (or lack of a threshold) for BP+OSD for a family of quantum LDPC codes, as well as the development of improved post-processing techniques that use BP marginals.
Despite these challenges, BP+OSD still has good performance for the large finite system sizes of 3D and 4D toric codes considered in this work, and so the asymptotic regime may not be so important in practice.

In future work, it would be interesting to investigate the performance of these higher dimensional hypergraph product codes using more realistic, circuit-level noise models.
We might expect the threshold of the 4D toric code to be competitive with that of the 2D surface code, given that we have observed a higher threshold here under a phenomenological noise model, and the 4D toric code only has a slightly higher check-weight of six, relative to the weight-four checks of the surface code.
The circuit-level noise can be handled by using BP+OSD to decode the full circuit-level Tanner graph~\cite{higgott2022fragile}, in which there is a variable node for each possible fault mechanism in the stabiliser measurement circuit.
However, this requires the use of much larger check matrices, so more efficient post-processing methods than OSD will likely be required.
Finally, additional improvements in performance can be expected by using soft information to update the priors on measurement errors~\cite{pattison2021improved,raveendran2022soft}.

\begin{acknowledgments} 
OH acknowledges support from the Engineering and Physical Sciences Research Council [grant number EP/L015242/1]. N.P.B. acknowledges support through the EPSRC Prosperity Partnership in Quantum Software for Simulation and Modelling (Grant No.
EP/S005021/1).
We thank Dan Browne and Joschka Roffe for feedback on an earlier version of the manuscript.
\end{acknowledgments}
	
\bibliography{references}

\appendix

\section{Update rules based on log-likelihood ratios}\label{app:bp_update_rules}

In this section we will review variants of BP that we used in this work, however we refer the reader to Ref.~\cite{chen2005reduced} for a more complete overview of BP.
In \Cref{sec:bp} we reviewed the update rule known as the ``tanh rule'' for implementing BP when representing probabilities using log-likelihood ratios (LLR).
The log-likelihood ratio $L(U)$ of a binary random variable $U$ is defined as 
\begin{equation}
L(U)=\log\left(\frac{P(U=0)}{P(U=1)}\right)
\end{equation}
where here the natural logarithm is used. 
It follows that $P(U=0)=\frac{e^{L(U)}}{1+e^{L(U)}}$ and $P(U=1)=\frac{1}{1+e^{L(U)}}$. 
In this representation, the sum (modulo 2) of two independent binary random variables $U$ and $V$ is given by
\begin{align}
L(U\oplus V)&\coloneqq \log\left (\frac{P((U\oplus V)=0)}{P((U\oplus V)=1)}\right) \nonumber \\
&=\log\left(\frac{P(U=0)P(V=0)+P(U=1)P(V=1)}{P(U=0)P(V=1)+P(U=1)P(V=0)}\right) \nonumber\\
&=\log\left(\frac{e^{L(U)+L(V)}+1}{e^{L(U)}+e^{L(V)}}\right)\nonumber\\
&=2 \tanh ^{-1}\left(\tanh \left(\frac{L(U)}{2}\right) \tanh \left(\frac{L(V)}{2}\right)\right).\label{eq:tanh_rule_rewrite}
\end{align}
The identity in \Cref{eq:tanh_rule_rewrite} is the origin of the ``tanh rule'' horizontal update rule for implementing BP using LLRs (see \Cref{eq:horiz_tanh}).
However, using the tanh rule directly in an implementation of BP can lead to at least two problems.
Firstly, once the LLR messages become sufficiently large, then a direct implementation of the horizontal tanh rule updates as given in \Cref{eq:horiz_tanh} using floating point arithmetic is numerically unstable.
Secondly, the hyperbolic tangent computations can be too expensive for some hardware implementations. 

\begin{figure}
\centering
\includegraphics[width=0.8\columnwidth]{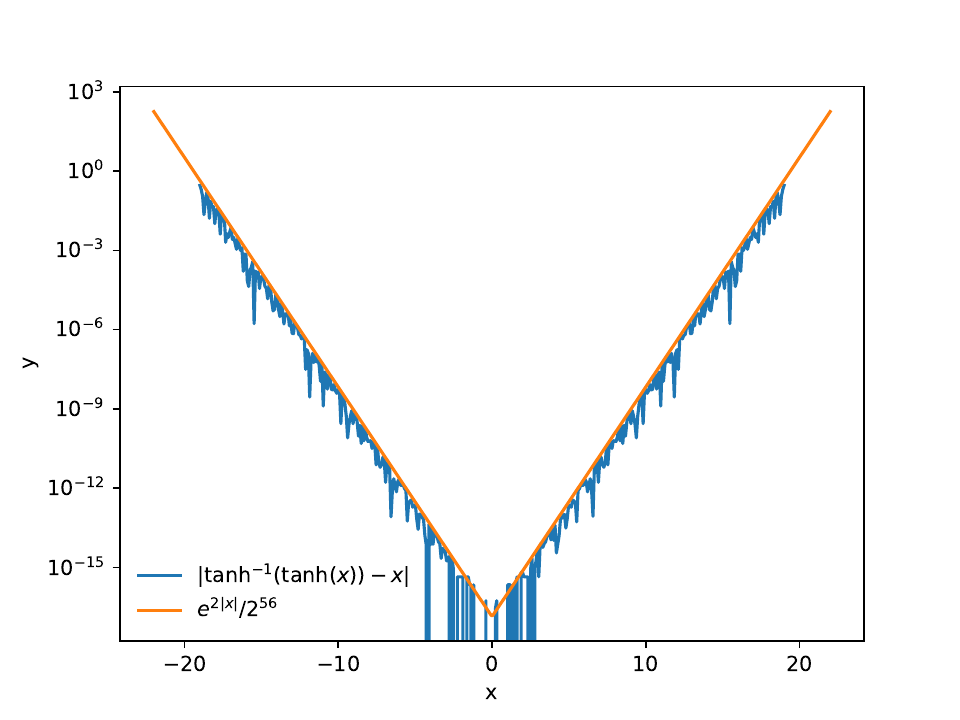}
\caption{The numerical error caused by underflow when evaluating $\tanh^{-1}(\tanh(x))$.}\label{fig:tanh_underflow}
\end{figure}

The numerical instability of the tanh rule can be understood by considering the simpler problem of evaluating
\begin{equation}\label{eq:arctanhtanh}
f(x)\coloneqq\tanh^{-1}(\tanh(x))=x
\end{equation}
using floating point arithmetic. Consider the regime $x\gg0$ where $\tanh(x)\approx 1$. In this regime, the round-off error when evaluating $\tanh(x)$ using IEEE double precision is approximately constant and equal to $\partial_{\tanh(x)}\approx 2^{-54}$. This results in an error in $f(x)$ of 
\begin{align}
\partial_{f(x)}&\approx\frac{df(x)}{d\tanh(x)}\partial_{\tanh(x)}\\
&\approx\frac{e^{2x}}{4} \partial_{\tanh(x)} \approx \frac{e^{2x}}{2^{56}}.
\end{align}
Since $\tanh$ and $\tanh^{-1}$ are both antisymmetric, the same holds for $x\ll 0$, i.e.~the numerical error scales exponentially in $|x|$, as $\partial_{f(x)}\approx e^{2|x|}/2^{56}$, as also shown empirically in Fig.~\ref{fig:tanh_underflow}.
One approach to handle these underflows is to cap the magnitude of the log-likelihood ratios when evaluating the check-to-variable messages. 
Alternatively, \Cref{eq:horiz_tanh} can be implemented exactly using the Jacobian approach, outlined in \Cref{sec:jacobian}, which does not suffer from these numerical stability issues.

However, the Jacobian approach requires evaluating logarithms and exponentials and, as with the tanh rule, is not so amenable to fast implementations in hardware.
The min-sum update rule is an \textit{approximation} of the tanh rule which avoids its numerical instability issues as well as requiring only real additions, simplifying its implementation.
We review the min-sum update rule in \Cref{sec:minsum}

\subsection{The Jacobian approach}\label{sec:jacobian}

The horizontal tanh rule of \Cref{eq:horiz_tanh} can be rewritten in a form that does not suffer from numerical underflow issues. 
To derive this alternative update rule~\cite{chen2005reduced, erfanian1994reduced}, first notice that
\begin{align}
e^a+e^b&=e^a(1+e^{b-a})\nonumber\\
&=e^{a+\log(1+e^{b-a})}\nonumber\\
&=e^{\max(a,b)+\log(1+e^{-|a-b|})}
\end{align}
from which we see that we can rewrite \Cref{eq:tanh_rule_rewrite} as
\begin{align}
L(U\oplus V)&=\log\left(\frac{e^{L(U)+L(V)}+1}{e^{L(U)}+e^{L(V)}}\right)\nonumber\\
&=L(V)+\max(L(U),-L(V))-\max(L(U),L(V))\nonumber\\
&\qquad+\log(1+e^{-|a+b|})-\log(1+e^{-|a-b|})\nonumber\\
&=\mathrm{sign}(L(U))\mathrm{sign}(L(V))\min(|L(U)|,|L(V)|)\nonumber\\
&\qquad+\log(1+e^{-|a+b|})-\log(1+e^{-|a-b|})\label{eq:jacobian_rule}.
\end{align}

In order to implement the log-likelihood horizontal update rule using \Cref{eq:jacobian_rule}~\cite{chen2005reduced}, we first assume that the factor $f_j$ is connected to $k$ variable nodes on the factor graph labelled $v_1, v_2,\ldots,v_k$, (i.e. we assume $\partial j=\{1,2,\ldots,k\}$). We now define $f_1=v_1, f_2=f_1\oplus v_2, \ldots,f_k=f_{k-1}\oplus v_k$ and $g_k=v_k, g_{k-1}=g_k\oplus v_{k-1}, \ldots,g_1=g_2\oplus v_1$, and then calculate all $L(f_i)$ and $L(g_i)$ using \Cref{eq:jacobian_rule} recursively (e.g.~$L(f_i)=L(f_{i-1}\oplus v_i)$ and $L(g_i)=L(g_{i+1}\oplus v_i)$). Now since $v_1\oplus v_2 \oplus \ldots \oplus v_k=z_j$, we can write $v_i=z_j\oplus f_{i-1}\oplus g_{i+1}$, from which we see that $L(v_i)=(-1)^{z_j}L(f_{i-1}\oplus g_{i+1})$. Therefore, the horizontal messages are computed using this `Jacobian rule' for $1<i<k$ as
\begin{equation}
Q_{f_j\rightarrow v_i}=(-1)^{z_j}L(f_{i-1}\oplus g_{i+1}),
\end{equation}
and the remaining messages are calculated as $Q_{f_j\rightarrow v_0}=(-1)^{z_j}g_1$ and $Q_{f_j\rightarrow v_k}=(-1)^{z_j}f_{k-1}$.

\subsection{The min-sum rule}\label{sec:minsum}

While the tanh rule and Jacobian rule can both be implemented efficiently in a time linear in the blocklength, for some practical applications it becomes necessary to use an even simpler decoder, which avoids the costly evaluation of the hyperbolic tangent. A widely used update rule is the min-sum rule, which approximates the tanh rule while being significantly simpler to implement in hardware~\cite{fossorier1999reduced,chen2005reduced}. From \Cref{eq:jacobian_rule} we see that $|L(U\oplus V)|\leq \min(|L(U)|,|L(V)|)$, which leads us to the min-sum horizontal update rule
\begin{equation}
Q_{f_j\rightarrow v_i}=(-1)^{z_j}\alpha\min_{i^{\prime}\in\partial j\setminus i}(Q_{v_{i^\prime}\rightarrow f_j})\prod_{i^{\prime}\in\partial j\setminus i}\mathrm{sign}(Q_{v_{i^\prime}\rightarrow f_j})
\end{equation}
where here $0\leq\alpha\leq 1$ is a constant chosen to better approximate the tanh rule. 
By setting $\alpha=2^{-a}+2^{-b}$ for some $a,b\in\{1,2,3\}$, multiplications can be replaced with bit shifts and additions, which simplifies FPGA implementations~\cite{valls2021syndrome}.

\section{Locality of BP}\label{sec:bp_locality}

\begin{figure}
    \centering
    \includegraphics[width=0.95\columnwidth]{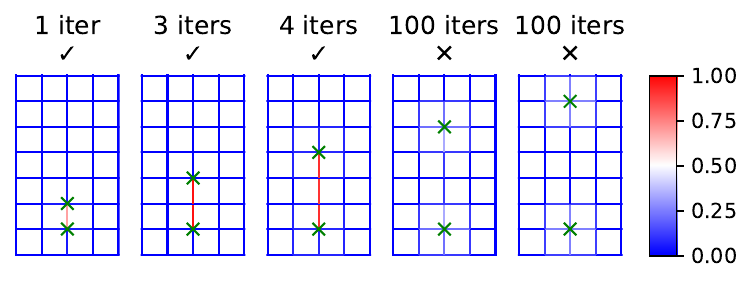}
    \caption{The marginals output by BP for an $L=15$ toric code for five different syndromes (and with a prior of $0.05$ for each variable node in the Tanner graph). We consider the problem of decoding $Z$ errors using $X$ stabiliser measurements, and a qubit is associated with each edge and an X stabiliser with each vertex of the lattice. For each syndrome we only show a $4\times 7$ region of the lattice. A -1 $X$ stabiliser measurement is displayed as a green cross, and the colour of each edge shows the marginal probability output by BP (according to the colour bar on the right). The title for each example syndrome shows the number of iterations used by BP, as well as a tick if BP converged or a cross if it did not converge. We use a maximum number of iterations of 100.}
    \label{fig:bp_locality}
\end{figure}

We observed in \Cref{sec:limitations_of_bp_osd} that increasing the maximum number of iterations of BP beyond some fixed number did not noticeably impact performance, even for large distance codes.
This suggests that it was not the number of iterations that was limiting performance for the 2D toric code.
To understand this better, in \Cref{fig:bp_locality} we consider the simple problem of decoding a single vertical string-like error in the 2D toric code using BP.
In other words, the syndrome is a pair of -1 $X$ stabiliser measurements, separated by some number $l$ of vertical edges in a vertical column of the lattice.
We will refer to the locations of -1 stabiliser measurements as \textit{defects}.
We have deliberately set up the problem in such a way that there is only a single minimum-weight solution, to reduce the impact of ``split-beliefs'' due to degenerate solutions in quantum codes.

We find that, for $l\leq 3$, BP converges quickly, as shown by the marginals in the left three examples in \Cref{fig:bp_locality}.
However, when the defects are separated by 4 or more edges (e.g.~$l=4$ and $l=5$ for the right two examples in \Cref{fig:bp_locality}), BP instead fails to converge.
This suggests that information is only propagated effectively by BP within some local region of the lattice.
Since BP is exact on tree graphs, and split beliefs are not a problem for this example, we expect that this issue arises due to loops in the Tanner graph.
For the 2D toric code, the shortest loop in the Tanner graph for $Z$ errors (e.g.~considering only the $X$ check matrix) has length 8; for each length-4 loop around a face of the lattice, there is a corresponding length-8 loop in the Tanner graph.
Similarly, the defects separated by $l=4$ edges in the lattice are in fact also separated by $8$ edges in the Tanner graph.
This suggests that the ability of BP to effectively propagate information significantly degrades beyond a radius in the Tanner graph that corresponds to its girth (some information propagates, but the strength of the signal is reduced).
We refer to this as the problem of \textit{bounded information spread} in BP.
Note that we are not claiming that \textit{no} information propagates beyond the radius equal to the girth.
Indeed, we verified that there is still enough information in the BP marginals for $l=4$ and $l=5$ for OSD to find the minimum-weight correction in \Cref{fig:bp_locality}, since the marginal probabilities are slightly larger along the minimum-weight path (albeit much less than 0.5).
However, in a more realistic setting in which more defects are present, we might expect the messages propagated from nearby defects (within the radius equal to the girth) to overwhelm or `hide' the much weaker messages passed from further away in the Tanner graph.
Therefore, there may be limited benefit in increasing the maximum number of iterations of BP with system size for codes with Tanner graphs that have constant girth, and so it is reasonable to leave this parameter as a constant (30 in our case).
Note that the full Tanner graph of a quantum code in general has girth 4, in contrast to the Tanner graph of either the $X$ check matrix or $Z$ check matrix alone when decoding $X$ and $Z$ errors independently, as we have done here, which potentially worsens the problem of bounded information spread in BP.
This is due to the commutativity condition of quantum codes.
For example, for the full Tanner graph of a CSS quantum code, each $X$ stabiliser must overlap on an even number of qubits with each $Z$ stabiliser.
Consider an $X$ stabiliser $S_X$ and $Z$ stabiliser $S_Z$ which overlap on two qubits~$i$ and~$j$.
In the full Tanner graph there will be variable nodes corresponding to $Y$ errors $Y_i$ and $Y_j$.
As a result there will be a loop (containing four edges) in the full Tanner graph that visits nodes in the order $(S_X,Y_i,S_Z,Y_j,S_X)$.

Our analysis does, however, further emphasises the limitations of BP alone for the 2D toric code, and so post-processing is crucial to ensure that these string-like errors for $l\geq 4$ can be corrected.
While the inference performed by BP is useful, it appears important (for the 2D toric code at least) that the post-processing step be performed by a decoder that considers global syndrome information.
For quantum codes where these issues of split beliefs and locality arise, a natural strategy might be to take some decoder known to have a threshold for the code of interest, and then use BP marginals to boost its performance.
That way we do not rely on the BP marginals alone to provide a correction, since these marginals may not be sufficiently reliable.

\section{Additional numerical results for balanced product codes}\label{sec:balanced_product_vary_noisy_rounds}

\begin{figure}
\includegraphics[width=0.8\columnwidth]{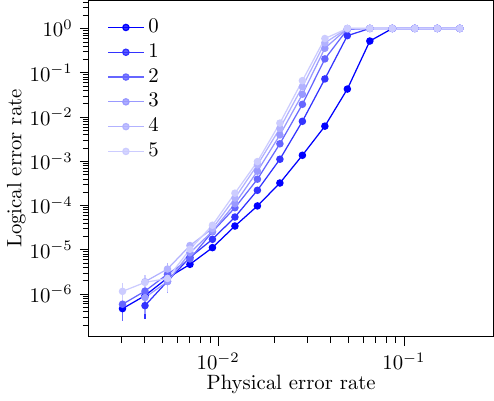}
\caption{Single shot performance of the [[1628,100]] balanced product code as the number of rounds of noisy measurements (given in the legend) are varied.}
\label{fig:balanced_product_vary_rounds}
\end{figure}

In \Cref{fig:balanced_product_vary_rounds} we show the single-shot decoding performance of the [[1628,100]] balanced product code for a varying number of rounds $r$ of noisy syndrome measurement (before one final round of perfect syndrome measurement).
In other words, $r=0$ corresponds to perfect syndrome measurement (code capacity model).
The logical error rate increases with the number of rounds as expected, but the increase in logical error rate remains stable, increasing approximately linearly with the number of rounds for $r>1$.

\end{document}